\documentclass[aps,prl,twocolumn,groupedaddress,floatfix]{revtex4}

\usepackage{graphicx}
\usepackage{bm}

\begin{document}


\title{Random Fields and the Partially-Paramagnetic State of 
CsCo$_{0.83}$Mg$_{0.17}$Br$_3$: \\A Critical Scattering Study}


\author{J. van Duijn$^{1}$}
\author{B.D. Gaulin$^{1,3}$}
\author{M.A. Lumsden$^{1}$}
\author{W.J.L. Buyers$^{2,3}$}
\affiliation{$^1$Department of Physics and Astronomy, McMaster University, 
Hamilton, Ontario, L8S 4M1, Canada}
\affiliation{$^2$Neutron Program for Materials Research, National Research 
Council, Chalk River Laboratories, Chalk River, K0J 1J0, Canada}
\affiliation{$^3$Canadian Institute for Advanced Research, 180 Dundas 
Street W, Toronto, ON, M5G 1Z8, Canada}

\date{\today}

\begin{abstract} Critical neutron scattering measurements have been
performed on CsCo$_{0.83}$Mg$_{0.17}$Br$_3$, a dilute stacked triangular
lattice (STL) Ising antiferromagnet (AF). A two component lineshape
associated with the critical fluctuations appears at a temperature
coincident with T$_{N1}$ observed in pure CsCoBr$_3$.  Such scattering is
indicative of fluctuations in prototypical random field Ising model (RFIM)
systems. The random field domain state arises in this case due to
geometrical frustration within the STL Ising AF, which gives rise to a 3
sublattice Neel state, in which one sublattice is disordered.  Magnetic
vacancies nucleate AF domains in which the vacancies reside on the
disordered sublattice thereby generating a RFIM state in the absence of an
applied magnetic field.

\end{abstract}


\maketitle


Geometrically frustrated magnetic materials have been of intense recent
interest, in part due to the diversity in the exotic ground states that 
they
display~\cite{frustration}.  These span the range from complex
non-collinear Neel states, to disordered co-operative paramagnetic and 
spin
liquid states, to spin glass-like states in the presence of little or no
chemical disorder.  Among the exotic magnetic states known to exist are
partially-paramagnetic Neel states, characterized as multi-sublattice
antiferromagnets, in which one of the sublattices in this long
range ordered structure remains paramagnetic. This unusual structure 
occurs in
several magnetic materials made up of triangular planes, including the
stacked triangular lattice Ising antiferromagnetic insulators 
CsCoBr$_3$~\cite{CsCoBr3}
and CsCoCl$_3$~\cite{CsCoCl3}, as well as the hexagonal intermetallic 
compound UNi$_4$B~\cite{UNi4B}.

A striking feature of the phase behavior of CsCoBr$_3$ and CsCoCl$_3$ is
their extreme sensitivity to the presence of non-magnetic impurities
on the magnetic Co site.  Neutron scattering work on
CsCo$_{1-x}$Mg$_x$Cl$_3$ by Nagler {\it et al.}~\cite{CsCoMgCl3} and on 
CsCo$_{1-x}$Mg$_x$Br$_3$ by Rogge {\it et al.}~\cite{CsCoMgBr3} show that 
17$\%$ dilution of 
the cobalt sites with non-magnetic magnesium appears to wipe out at least 
two (and possibily three) long range ordered states which pure 
CsCoCl$_3$ displays below T$_{N1}$$\sim$ 21 K~\cite{CsCoCl3}, and which 
pure CsCoBr$_3$ 
displays below $\sim$ 28 K~\cite{CsCoBr3}.  This is surprising as 
3 dimensional lattices support percolating long 
range order to $\sim$ 75$\%$ dilution.~\cite{Chaikin}. 

In this letter, we report on a critical neutron scattering study
of single crystal CsCo$_{0.83}$Mg$_{0.17}$Br$_3$.  We show that the
critical fluctuations contain two components, the narrower of which (in
$\vec{Q}$-space) displays an onset at a temperature co-incident with
T$_{N1}$ of pure CsCoBr$_3$, $\sim$ 28 K.  Two distinct components to the
fluctuations are easily observed over the approximate temperature interval
for which pure CsCoBr$_3$ exhibits the exotic partially-paramagnetic,
3-sublattice Neel state.  We argue that this two component
lineshape is characteristic of a domain state typical of the Random Field
Ising Model (RFIM), and which is generated by the presence of quenched 
magnetic
vacancies and the paramagnetic sublattice.

RFIMs occupy interesting and important territory within the physics of
disordered media~\cite{RFIM}. Within this model the disorder arises due to 
random
local symmetry breaking, as opposed to random interactions and
frustration, which drives the physics of conventional spin 
glasses~\cite{spinglass}.  The
RFIM is known to describe well the properties of magnetically-dilute Ising
antiferromagnets in the presence of magnetic 
fields~\cite{coznf2,mnznf2,feznf2,rbcomgf4}, as well as a wide
class of other disordered materials, such as, for example, phase 
separating fluids in the presence of gels~\cite{gels}.

Imry and Ma first considered the effects of a ferromagnetic Ising system
in the presence of a site-random magnetic field, and argued that the
random fields would destroy the long range order of such a classical
system for dimensions less than 4~\cite{imry}.  Fishman and Aharony later
argued that magnetically dilute Ising antiferromagnets in the presence of
a uniform magnetic field were isomorphous to the Imry-Ma
scenario~\cite{fishman}. This was important as not only do
site-random antiferromagnets occur in nature, but it is easy 
to apply
an external magnetic field and thereby tune the amplitude of the random
field.

Systematic studies of such site-random Ising antiferromagnets have
been carried out for both three dimensional materials (such as
CoZn$_{1-x}$F$_2$~\cite{coznf2}, 
MnZn$_{1-x}$F$_2$~\cite{mnznf2}, 
and FeZn$_{1-x}$F$_2$~\cite{feznf2}) as well as
quasi-two dimensional systems (such as 
Rb$_2$Co$_{0.7}$Mg$_{0.3}$F$_4$~\cite{rbcomgf4}).  
While some open questions remain,
it is clear that the field-cooled state at low temperatures corresponds to
a system broken up into many relatively small domains.  A characteristic
signature of such a state in neutron scattering experiments is a 
two-component lineshape in wave vector, with a sharp and a broad
component, as opposed to the single-component lineshape
corresponding to, say, Ornstein-Zernike fluctuations~\cite{review}.  Such
a two-component lineshape, however, was inferred from the results of 
fitting the neutron lineshapes to an appropriate addition of two terms.

\begin{figure}
\includegraphics{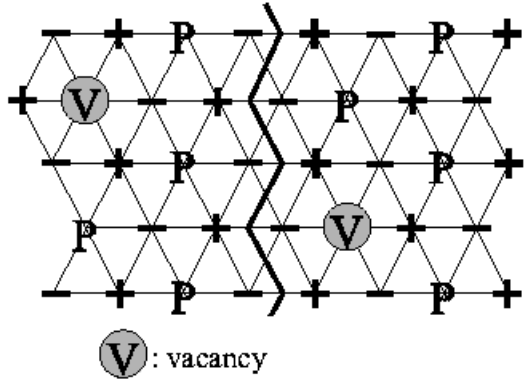}
\caption{Role of non-magnetic atoms in generating a random
field domain state in CsCo$_{0.83}$Mg$_{0.17}$Br$_3$}\label{fig2}
\end{figure}

CsCo$_{0.83}$Mg$_{0.17}$Br$_3$ belongs to the ABX$_3$ family of
antiferromagnetic (AF) insulators~\cite{abx3}. The interest in 
these materials stems from the quasi-one-dimensional nature of 
their magnetic
interactions, as well as their geometrically frustrated cooperative
behaviour. Pure CsCoBr$_3$ is an example of an Ising-like 
antiferromagnet
on a stacked triangular lattice~\cite{CsCoBr3}. It crystallizes in the
hexagonal space group P$6_3$/\textit{mmc}, with low temperature lattice
parameters a= 7.45 and c= 6.30 \AA. The Co$^{2+}$ ions possess
spin-$\frac{1}{2}$ magnetic moments, which are aligned 
along the hexagonal \textbf{c} direction, and lie on 
a simple hexagonal lattice.  Inelastic neutron scattering studies have shown
that this material is described by the following spin
Hamiltonian~\cite{nagler}: 
\begin{eqnarray} \mathcal{H} & = &
\sum_{i(\vec{c})}(2J\{S_i^zS_{i+1}^z + \epsilon (S_i^xS_{i+1}^x +
S_i^yS_{i+1}^y)\} \nonumber\\
 &  & + h_0S_i^z(-1)^i + 2J'\sum_{j(ab)}^6S_i^zS_{j}^z)
\end{eqnarray}
where \textit{J}= 1.62 THz, \textit{J'}= 0.0096 THz and $\epsilon$=
0.137. The dominant AF interactions along the \textbf{c}-axis
result in
quasi-one dimensional behaviour over an extended temperature
regime. At sufficiently low temperatures, however, the weaker
\textbf{ab}-plane interactions cause two or possibly three,
magnetic phase transitions to three dimensional long range ordered Neel
states~\cite{CsCoBr3,farkas}. The first phase transition,
from the paramagnetic state to a long range ordered, but 
partially-paramagnetic state, occurs at T$_{N1}$=28.3 K. 
This state is characterized by three sublattices, in which two out of 
every three spins on triangular plaquettes are ordered
antiferromagnetically, while the third remains disordered. This is 
shown as +, - and P in Fig~\ref{fig2}. The ordering triples the 
periodicity within the
\textbf{ab}-plane, giving rise to magnetic Bragg reflections of
the form ($\frac{h}{3},\frac{h}{3},l$) with l odd. Below $\sim$ 13 K the
paramagnetic sublattice orders fully, either (+) up or down (-), to form 
a ferrimagnetic sheet within the \textbf{ab}-plane.  The net moment within 
this ferrimagnetic sheet orders alternately from \textbf{ab}-plane to
\textbf{ab}-plane, so as to produce an antiferromagnetic structure, with 
no net magnetic moment. An intermediate magnetically-ordered structure may 
exist between $\sim$ 18 K and 13 K, in which the paramagnetic sublattice 
gradually orders.  

Classical Ising antiferomagnets on stacked triangular lattices have been 
studied extensively by Landau-Ginsburg-Wilson~\cite{landau} and Monte 
Carlo techniques~\cite{montecarlo}.
The high and low temperature long range ordered states are accounted for 
by all theoretical models.  
\begin{figure} \includegraphics{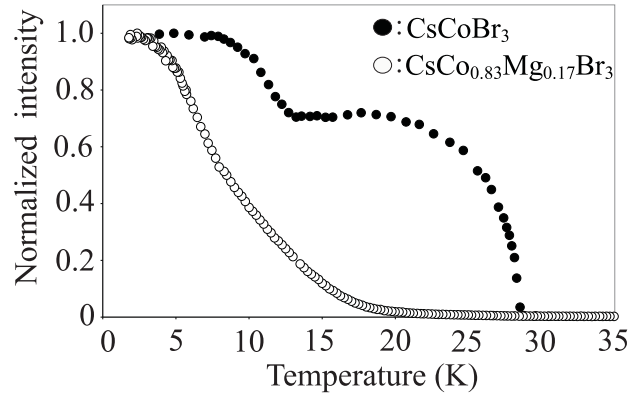}
\caption{Peak intensity of the magnetic
($\frac{2}{3},\frac{2}{3},1$) and ($\frac{4}{3},\frac{4}{3},1$)
Bragg reflection of CsCoBr$_3$ and CsCo$_{0.83}$Mg$_{0.17}$Br$_3$
respectively as a function of temperature.}\label{fig1}
\end{figure}

The argument by which the combination of a partially-paramagnetic Neel 
state and magnetic vacancies generate a random field domain state is 
illustrated in Fig. 1. These quenched vacancies will preferentially 
nucleate domains in which the vacancies are coincident with the 
paramagnetic sublattice so as to minimize the loss 
of exchange energy. Two such 
domains are depicted in Fig. 1, with a domain wall running 
between them.  In the absence of the paramagnetic sublattice, the 
vacancies are co-incident with either up or down sublattices, and only 
break symmetry between the two in the presence of an externally applied 
longitudinal magnetic field; the Fishman-Aharony scenario~\cite{fishman}.

The critical neutron scattering study was carried out on the N5 triple
axis spectrometer, operated in two axis mode, at the NRU reactor of Chalk
River Laboratories. The high quality single crystal was the one 
studied by Rogge {\it et al.}~\cite{CsCoMgBr3}. Neutrons
of wavelength 2.37 $\AA$ were reflected from the $(0,0,2)$ 
planes of a
pyrolitic graphite (PG) monochromator. A PG filter was placed between 
the monochromator and sample to remove higher order contamination. The
sample, mounted in a He$^4$ cryostat, was oriented with its $(h,h,l)$
plane in the horizontal scattering plane. Collimation of 0.19$^\circ$ in
the incident beam and 0.25$^\circ$ in the diffracted beam was used to 
achieve
relatively high $\vec{Q}$ resolution.  

The temperature dependence of the Bragg peak intensities in both pure 
CsCoBr$_3$ and dilute CsCo$_{0.83}$Mg$_{0.17}$Br$_3$ is shown in 
Fig.~\ref{fig1}.  The square of the order parameters were taken 
to be proportional to the peak intensities at the magnetic 
Bragg reflection, $(\frac{2}{3},\frac{2}{3},1)$ for CsCoBr$_3$ 
and $(\frac{4}{3},\frac{4}{3},1)$ for CsCo$_{0.83}$Mg$_{0.17}$Br$_3$.

While more complete than that reported by Rogge {\it et
al.}~\cite{CsCoMgBr3}, the CsCo$_{0.83}$Mg$_{0.17}$Br$_3$ order parameter
squared presented in Fig. 2 reproduces the key characteristics of the
earlier measurements.  The Bragg intensity displays upwards curvature down
to at least 5 K, and the clear phase transitions in pure CsCoBr$_3$ at
$\sim$ 28.3 K and $\sim$ 13 K are not evident.  Mean field behavior would
predict this intensity to grow linearly with decreasing temperature as
$\sim$ (T$_C$-T)$^{2\beta}$ with $\beta$=0.5.  Typical three dimensional
critical behavior would give $\beta$$\sim \frac{1}{3}$. Pronounced
downwards curvature in the Bragg intensity as a function of temperature
would be expected, and indeed is observed in pure CsCoBr$_3$ as shown in
Fig. 2.

Similar order parameter measurements on CsCo$_{0.83}$Mg$_{0.17}$Cl$_3$ by
Nagler {\it et al.}~\cite{CsCoMgCl3} also show upwards curvature to the 
Bragg 
intensity as a function of temperature to temperatures below 5 K,
and again no evidence of the T$_{N1}$$\sim$ 21 K observed in pure 
CsCoCl$_3$.  While these earlier studies concluded phase 
transitions occur near 9 K and 6 K, for CsCo$_{0.83}$Mg$_{0.17}$Br$_3$ and 
CsCo$_{0.83}$Mg$_{0.17}$Cl$_3$ respectively, neither study showed 
compelling evidence for any phase transition above $\sim$ 5 K.

\begin{figure} \includegraphics{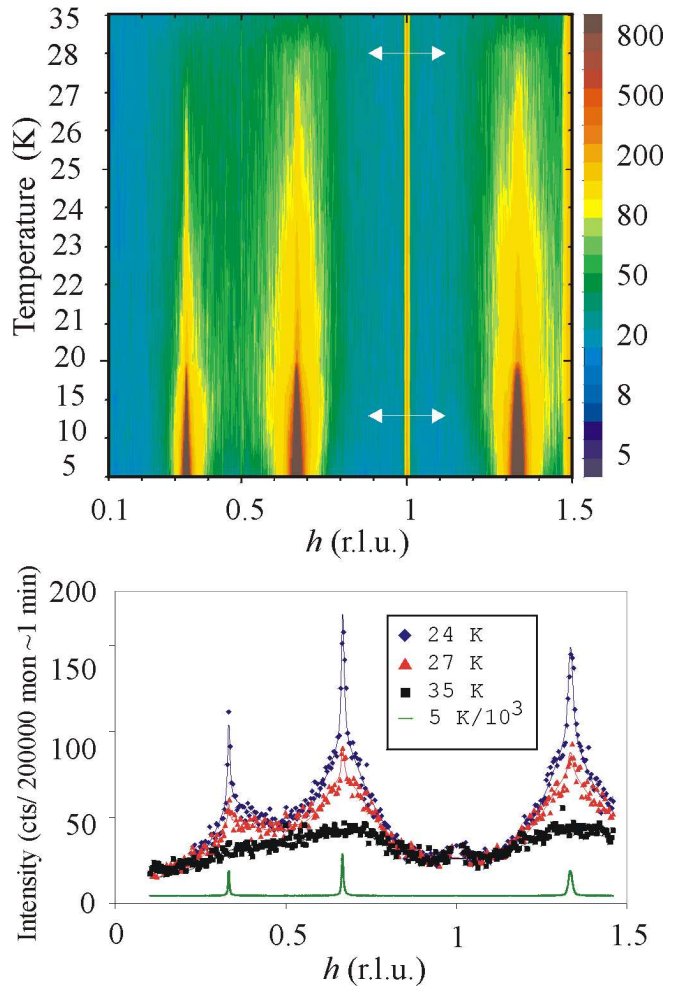} \caption{
(top) Scattering data along the $(h,h,1)$ direction at the
13 measured temperatures. The white arrows indicate the temperatures
at which clear phase transitions are observed in pure CsCoBr$_3$ (see Fig. 
2).  (bottom) Representative $(h,h,1)$ scans clearly showing the onset of 
a two-component line shape below T$_{N1}$. The solid
lines are fits of the data to Eq. 2, convoluted with the experimental
resolution. The bottom data set, scaled in intensity 
by a factor of 10$^3$, shows that only the sharp component remains at 5 K.}
\label{fig3} \end{figure}

As a function of temperature, scans in $h$ and $l$ were performed through
three ordering wavevectors: 
$\vec{Q}_{ord}$=
$(\frac{1}{3},\frac{1}{3},1)$,$(\frac{2}{3},\frac{2}{3},1)$ and
$(\frac{4}{3},\frac{4}{3},1)$. 
As shown by the h-scans in Fig.~\ref{fig3} a
sharp, albeit relatively weak component, additional to the
broad critical fluctuations, appears below $\sim$ 28 K. This indicates 
that a
broken symmetry has occurred very close to pure CsCoBr$_3$'s T$_{N1}$.  
Over the temperature range $\sim$ 28 K to $\sim$ 20 K, the scattering
clearly displays a two component line shape. 

The data were analysed by convoluting the scattering cross section with
the measured 3-dimensional resolution function.
The cross section, centered at each ordering wavevector, is described by 
the sum of a broad and a narrow
Lorentzian, both having anisotropic widths.  The calculated intensities 
included both the Co$^{2+}$ magnetic form factor, and 
the neutron polarization factor, by which only components of 
moment perpendicular to $\vec{Q}$ contribute to the scattering.

The inverse correlation lengths within the \textbf{ab}-plane and along the
\textbf{c}-axis were allowed to differ, acounting for the strong
anisotropic nature of the AF interactions within this material; hence:
\begin{equation}
S(\vec{Q})=\frac{A}{1+\frac{q_a^2+q_b^2}
{\kappa_{ab}^{2}}+\frac{q_c^2}{\kappa_c^2}} + \frac{B}
{1+\frac{q_a^2+q_b^2}{\kappa_{ab}^{\prime2}}+\frac{q_c^2}{\kappa_c^{\prime2}}}
\end{equation} where $\kappa_{ab}<\kappa'_{ab}$, $\kappa_c<\kappa'_c$, and
$\vec{q}= \vec{Q}-\vec{Q}_{ord}$. The inverse correlation lengths of the
broad Lorentzian were allowed to refine, while those of the narrow 
one were held fixed (at $\kappa_{ab}$=0.00546 and $\kappa_c$=0.00624 
$\AA^{-1}$,
approximating their resolution limit). A complete fit to the data 
required adding two broad and relatively weak Lorentzians centered at 
wavevectors $(\frac{1}{2},\frac{1}{2},1)$ and
$(\frac{3}{2},\frac{3}{2},1)$, in addition to a constant background. This
extra scattering originates from the domain walls within the
\textbf{ab}-plane. A stable paramagnetic sublattice requires a net zero
exchange field from its near neighbours (3 up (+) and 3 down (-) near
neighbours).  The domain wall regions therefore exclude the paramagnetic
sublattice and reflect a local two sublattice structure (Fig.~\ref{fig2}),
giving rise to scattering at $(\frac{h}{2},\frac{h}{2},l)$ with $l$ odd.

\begin{figure}
\includegraphics{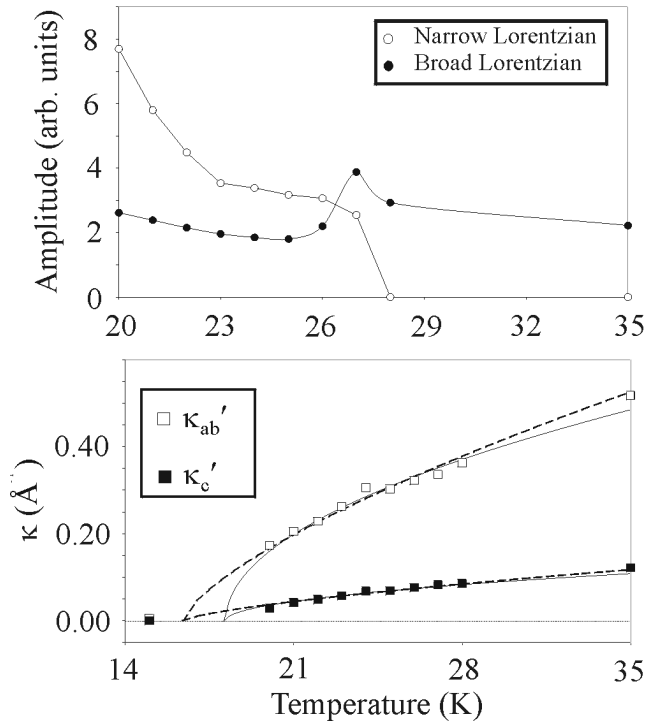}
\caption{(top) Temperature dependence of the amplitude of both the narrow 
and 
broad Lorentzians, resulting from fits shown in the bottom panel of Fig. 
3. (Bottom), The temperature dependence of 
the inverse correlation lengths of the broad Lorentzian are shown 
along with fits to divergences in the correlation lengths, appropriate to 
both mean field theory ($\nu$=$\frac{1}{2}$) and the 3D Heisenberg model
($\nu$=0.707), as described by Eq. 3}. 
\label{fig4} \end{figure}

Figure~\ref{fig4} shows the temperature dependence of the refined 
parameters.
The narrow component to the lineshape turns on below 28 K, while the broad
component extents above T$_{N1}$.  
Below 20 K the intensity of the narrow component rapidly increases, 
and the data can be equally well described by a single 
component Lorentzian with anisotropic line widths. 

To seek out a second, lower phase transition, we have fit the 
temperature dependence of the inverse 
correlation lengths of the broad 
component between 20 K and 28 K.  This was fit to a standard 
critical divergence of the 
correlation lengths as: 
\begin{equation} \kappa'_{ab,c} \sim 
\frac{1}{\xi'_{ab,c}} \sim
(T-T_{N2})^{\nu} \end{equation} 
We fit with critical exponents $\nu$ ranging from a low of $\frac{1}{2}$, 
appropriate to mean field theory, to a high of 0.707 appropriate to
the 3D Heisenberg model.  These fits, shown in the bottom panel of Fig. 4, 
are consistent with a second phase transition occuring between 
$\sim$ 16 K and 18 K.   

While a unique, quantitative analysis of the scattering below 20 K is 
difficult, examination of the qualitative features of the scattering in 
Fig. 3 is useful.  The order parameter for  
CsCo$_{0.83}$Mg$_{0.17}$Br$_3$ in Fig. 2 shows little or no indication of 
the $\sim$ 13 K transition in CsCoBr$_3$.  However the (h,h,1) scans in 
the 
top panel of Fig. 3 clearly show a narrowing in the diffuse scattering 
below 13 K, indicating the onset of a true long range ordered state.  We 
interpret this as a tendency towards full ordering of the paramagnetic 
sublattice below T$_{N3}$ $\sim$ 13 K, thereby removing the driving force 
for the RFIM behavior in the absence of an applied magentic field, that 
is the paramagnetic sublattice.

Our critical scattering study of 
CsCo$_{0.83}$Mg$_{0.17}$Br$_3$ 
has resolved a longstanding mystery~\cite{CsCoMgBr3, CsCoMgCl3} as to the 
extreme sensitivity of 
the long range ordered partially-paramagnetic Neel state to 
quenched magnetic vacancies.  It also contrasts with very recent 
theoretical work on doped quasi-1D Heisenberg 
antiferromagnets~\cite{Affleck}.  Our 
results show that  CsCo$_{0.83}$Mg$_{0.17}$Br$_3$ still displays a 
symmetry breaking near pure CsCoBr$_3$'s T$_{N1}$$\sim$ 28 K, but 
that the system enters a RFIM domain state, and remains in such a state until 
sufficiently low temperatures that the paramagnetic sublattice orders,
at approximately 13 K.  

This work was supported by NSERC of Canada.

\end{document}